\documentclass[12pt,a4paper]{article}
\usepackage{graphicx}
\usepackage{amsmath}
\usepackage{hyperref}
\usepackage{pstricks}
\usepackage{verbatim}
\usepackage[utf8x]{inputenc}
\hypersetup{%
  bookmarksopen=true,
  bookmarksnumbered=true,
  pdftitle = {The Grid[Way] Job Template Manager, a tool for parameter sweeping},
  pdfsubject = {Parameter sweep, Grid computing},
  pdfkeywords = {e-science, parameter sweep, grid computing, middleware, high-throughput computing},
  pdfauthor = {A. Lorca, E. Huedo, I.M. Llorente},
}
\def\linux{{\sc Linux}}

\newcommand{\eq}[1]{Eq.~(\ref{#1})}

\newcommand{\fig}[1]{Fig.~\ref{#1}}
\newcommand{\tab}[1]{Table \ref{#1}}
\newcommand{\tabs}[2]{Tables \ref{#1}$-$\ref{#2}}
\newcommand{\sect}[1]{Sect.~\ref{#1}}

\newenvironment{term}%
  {\endgraf\scriptsize\noindent\verbatim}%
  {\endverbatim}
 \title{The Grid[Way] Job Template Manager,\\ a tool for parameter sweeping}
 \author{\bigskip
{Alejandro Lorca\footnote{Corresponding author, e-Mail:\href{mailto:alejandro.lorca@fdi.ucm.es}{alejandro.lorca@fdi.ucm.es}}, Eduardo Huedo\footnote{e-Mail:\href{mailto:ehuedo@fdi.ucm.es}{ehuedo@fdi.ucm.es}}, Ignacio M. Llorente\footnote{e-Mail:\href{mailto:llorente@dacya.ucm.es}{llorente@dacya.ucm.es}}}\\* \vspace{-0.5ex}
{\small \em Facultad de Inform{\'a}tica, Universidad Complutense de Madrid,}\\*
{\small \em C/ Prof.\ Jos\'e Garc\'{\i}a Santesmases s/n, E-28040, Madrid, Spain}
}
\date{\today}
\begin{document}
\maketitle
\begin{abstract}
Parameter sweeping is a widely used algorithmic technique in computational science. It is specially suited for high-throughput computing since the jobs evaluating the parameter space are loosely coupled or independent. 

A tool that integrates the modeling of a parameter study with the control of jobs in a distributed architecture is presented. The main task is to facilitate the creation and deletion of job templates, which are the elements describing the jobs to be run. Extra functionality relies upon the GridWay Metascheduler, acting as the middleware layer for job submission and control. It supports interesting features like multi-dimensional sweeping space, wildcarding of parameters, functional evaluation of ranges, value-skipping and job template automatic indexation.

The use of this tool increases the reliability of the parameter sweep study thanks to the systematic bookkeping of job templates and respective job statuses. Furthermore, it simplifies the porting of the target application to the grid reducing the required amount of time and effort.
\end{abstract}
%
%
\section{Program summary}
\begin{itemize}
\item {\em Title of the program}: Grid[Way] Job Template Manager (version 1.0, released on February 26th, 2010).
\item {\em Catalogue identifier}:
\item {\em Program obtainable from}:\\ \url{http://dev.gridway.org/projects/gwjobtemplatemanager}.
\item {\em Computer}: any (tested on PC x86 and x86\_64).
\item {\em Operating system}: Unix, GNU/\linux{} (tested on Ubuntu 9.04, Scientific Linux 4.7, centOS 5.4), Mac OS X (tested on Snow Leopard 10.6).
\item {\em Programming language used}: perl 5.8.5 and above.
\item {\em Additional programs/libraries used}: The GridWay Metascheduler \cite{gridway}.
\item {\em Memory required to execute with typical data}: 10MB.
\item {\em No.~of processors used}: 1.
\item {\em No.~of bytes in distributed program, including test data, etc.}: 20799.
\item {\em Distribution format}: Gzipped tar file.
\item {\em High-speed storage required}: No.
\item {\em Keywords}: e-science, parameter sweep, grid computing, middleware, high-throughput computing.
\item {\em Nature of the physical problem}: To parameterize and manage an application running on a grid or cluster.
\item {\em Method of solution}: Generation of job templates as a cross product of the input parameter sets. Also management of the job template files including the job submission to the grid, control and information retrieval.
\item {\em Restriction of the complexity of the problem}: The parameter sweep is limited by disk space during generation of the job templates. The wildcarding of parameters cannot be done in decreasing order. Job submission, control and information is delegated to the GridWay Metascheduler.
\item {\em Typical running time}: From half a second in the simplest operation to a few minutes for thousands of exponential sampling parameters.
\end{itemize}
%
\section{Introduction}
Parameter sweeping is a very common strategy in order to solve complex scientific problems. It consists on scanning different points on a relevant parameter space region. There are many possible interests on doing such, for instance, the optimization of a functional by finding its minima over the scanned region, to parameterize an analytical function with the help of interpolated values or to understand different regimes of simulated physical models. The variety of the scientific disciplines profiting from this strategy is overwhelming; experimental high-energy physics, astrophysics and astro-particle physics, life sciences, computational chemistry,  earth sciences, computational linguistics, game theory, etc, are just prominent examples.

The evaluation of parameter space points is a simple problem within computational theory, but requires, in turn, some pre-processing of the problem according to a given search criteria, the systematic exploration of the parameter space points and a final data post-processing.

According to the application profile three different categories can be established for better understanding the scope of the tool:
\begin{itemize}
\item Parallelization. The second stage mentioned above, in which the parameter space is explored, requires the evaluation of some functional for each point. Making this stage {\em parallel} is in some cases unfeasible and the applications are run in {\em sequential} mode. Those parallelizable applications require some porting (i.e. either to cluster, grid or supercomputers) to improve the execution time, surmount memory limitations, etc.
\item Interactivity. User intervention might be automated to a certain degree depending on the application profile. If it is not required at any stage of the computation we will consider the applications as {\em non-interactive}. Otherwise, for the {\em interactive} ones, the user is obliged to take some decision during execution.
\item Recursion. On one hand, {\em single-pass} applications do not profit from the knowledge of the parameter space exploration or the results at a post-processing stage. On the other hand, if the application is aware of such feedback, an improved range of the parameter space can be chosen for optimizing results or further testing. This workflow is common in {\em master-worker} execution profiles.
\end{itemize}

With such classifications in mind, the {\em parallel, non-interactive, single-pass} applications are the target group for the Grid[Way] Job Template Manager, specially those with a large amount of independent tasks to be accomplished.

Large-scale problems well suited to high-throughput computing can be tackled in different manners, from last generation supercomputers to distributed solutions such as workstations or PCs. Indeed, grid computing has emerged during the last two decades as a new and affordable computational paradigm. The main distinguishing feature of this distributed architecture is the integration of different computing resources, typically involving high performance clusters (computing elements) and massive database storages (storage elements) hosted on universities, research centres, private companies, etc. The access to the grid is mediated through a software layer, usually referred to as the {\em middleware}, which takes care of the user's certificate checking, data transfer, host monitoring, file registering, and many other additional services.

There are other tools on the market dealing with parameter studies on the grid, like Nimrod/G \cite{Abramson:2000} or APST\cite{Casanova:2003} from AppLeS \cite{Casanova:2000}. Nimrod/G possess a ``parametric engine'' which is the agent centralizing the parameterization of the experiment, job creation, submission and control. Nimrod/G requires a static declarative language, allowing multi-dimensional analysis to set of strings and basic real ranges. Alternatively, the AppLeS Parameter Sweep Template uses a simple, XML-based\footnote{Extensible Markup Language.} interface which can be used from the command-line or called from scripts, but it is left to the user the enumeration of individual tasks. In both cases, the scheduling and other grid-related tasks are part of the projects. This is not the case of the Grid[Way] Job Template Manager, because the GridWay Metascheduler handles those operations directly.  As will be shown later, the reason for naming our tool with a bracket in the word Grid[Way] tries to make clear the relation to the grid technology, being optional the underlying use of GridWay.

GridWay itself contains a mechanism to proceed with a collection of jobs (i.e. job array)\cite{Huedo:2004}, but it is certainly limiting: only single parameter variation through ingeters, both in step and values, can be achieved. This is somewhat similar to the possibilities integrated on the gLite WMS job description language \cite{JDL}. A remarkable effort to stardardize the parameter sweeps has been carried out \cite{JSDL}, but the approach is more a formal extension of the job standard description language than a useful grammar easy to by remembered by the end user.

The structure of the paper is as follows: The \sect{algorithm} describes the method and framework used for the composition of the parameter sweep region directly from the spanning of the one-dimensional parameter sets. It also describes the syntax of the parameter file. \sect{setup} moves on to the technical steps to have the tool up and running while \sect{usage} shows the general use of the tool. Several examples have been worked out in detail in \sect{examples} with increasing dificulty. Finally we summarize the work in \sect{conclusion}.

\section{Algorithm}\label{algorithm}
\begin{figure}[!ht]
\center
\includegraphics[scale=0.5]{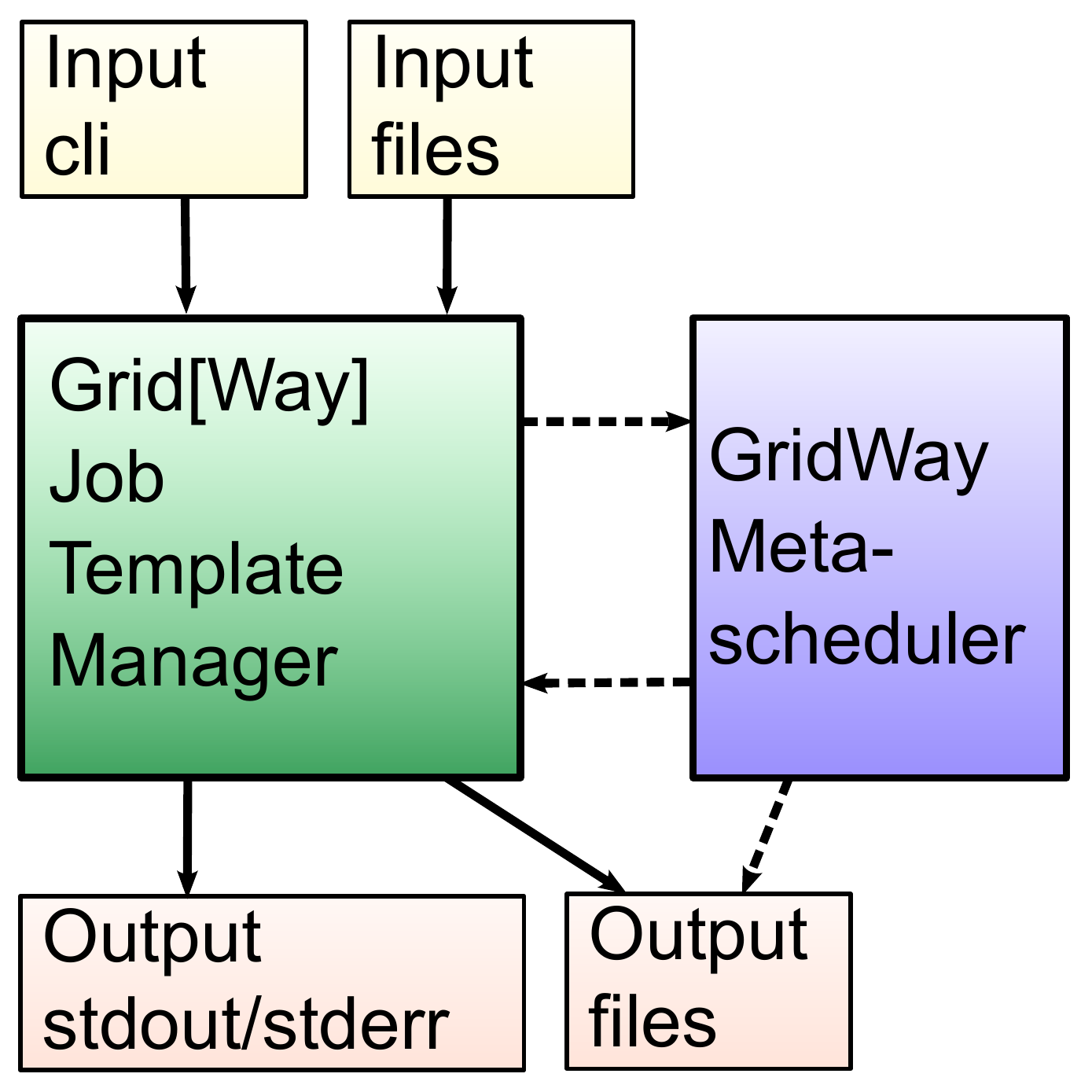}
\caption{Grid[Way] Job Template Manager framework. Continuous lines represent input/output while dashed lines indicates interaction with GridWay through optional instructions and information feedback.}
\label{basic}
\end{figure}

The basic working environment of the program is depicted in \fig{basic}. The user specifies what is going to be performed through the command line interface and, depending on which tasks has been asked for, some input files will be required. The parameter sweep specification comes from the parameter file and job template appendices from the template file. During the processing, some calls to the GridWay Metascheduler are required for submitting, purging, killing and getting information about jobs. The main outcome of the tool are the job template files, which contain the information about the parameter sweep study and are ready for submission. The information about the job creation and deletion, as well as from the jobs coming from these job template files is available through the standard output and error. Additionally, if GridWay is used, the corresponding jobs' output will be available as extra files.

\subsection{Mathematical framework}
The starting point for carrying out a parameter study is to define which parameter points are considered for evaluation. Let denote by ${\bf r}$ a parameter point belonging to the direct product space of suitable dimensions ${\bf R}$:
\begin{equation}
{\bf r} \in {\bf R}, \quad {\bf R}=R_1 \times \cdots \times R_n,
\end{equation}
being $R_i$ the respective one-dimensional space of the $i$-th argument.

A given set of points $P$ of arbitrary size $m$ forms the complete parameter sweep set:
\begin{equation}
P = \{ {\bf r}_j : {\bf r}_j \in {\bf R}; 0 \leq j \leq m-1 \},
\end{equation}
but for simplification we will consider only the case where any element ${\bf r}_j$ comes from a Cartesian product of subsets $P_i \subset R_i$ of some chosen elements, each subset with size $m_i$. Thus,
\begin{eqnarray}
P&=&P_1 \times P_2 \times \cdots \times P_n, \label{cartesian}\\
m&=&m_1m_2 \cdots m_n=\prod_{i=1}^n m_i. \label{size}
\end{eqnarray}

Indexing the $n$-tuple elements of $P$ is trivial when following the natural ordering of each set $P_i$ until it gets exhausted, then incrementing the previous set.

Note the difference in labeling convention: the elements in each set $P_i$ are indexed from $0$ to $m_i-1$, while the composition of the $n$-tuples correspond to $1,2,\ldots,n$. This is because we expect the user to feed the application with $n$ command line arguments, being each job uniquely indexed:
\begin{equation}\label{indexing}
\begin{split}
{\bf r}_0&=(r_{1,0}, r_{2,0}, \ldots , r_{n,0}),\\
{\bf r}_1&=(r_{1,0}, r_{2,0}, \ldots , r_{n,1}),\\
\vdots&=\vdots\\
{\bf r}_{m-1}&=(r_{1,m_1-1}, r_{2,m_2-1}, \ldots, r_{n,m_n-1}).
\end{split}
\end{equation}

We have not yet described which is the allowed nature of the spaces $R_i$ where the subsets $P_i$ are included. Let us discuss this issue altogether with the mechanism to model the study.

\subsection{Parameter file parsing}
Next, the way to declare the sets $P_i$ for each argument is explained. The declaration grammar has been chosen to be fairly simple for user's convenience:
\begin{itemize}
\item The parametric study, denoted by $P$ is declared as $n$ valid sentences in the parameter file.
\item Each set $P_i$ is declared at the valid $i$-th sentence of the file, one per line. It contains instructions to define the structure of the set and value assignments.
\item The words of the language are KEY=VALUE pairs, separated by blank spaces.
\end{itemize}

The structure of each set is declared in the first word. The following possibilities are currently available:
\begin{enumerate}
\item {\tt LOOPTYPE=LIST}. The set consists of a list of given values. These are assigned through an arbitrary number of {\tt VALUE=}$r_{i,j}$ pairs. Admitted values are taken as strings.
\item {\tt LOOPTYPE=RANGE}. The set is built according to a linear range of values. The keys {\tt START} and {\tt END} define the initial and final elements of the set, while the tags {\tt STEP} or {\tt POINTS} fix either the iteration increment between each element or the total number of points of the set respectively. Admitted values can be entered in the following formats:
\subitem a) integer (i.e. {\tt -4} or {\tt 54764563}),
\subitem b) floating point up to 16 digits (i.e. {\tt 567884.2234}),
\subitem c) scientific up to $10^{\pm 308}$ (i.e. {\tt 1.4E-12}),
\subitem d) characters (i.e. {\tt j} or {\tt z}),
\subitem e) arbitrary-size\footnote{Should the user require this feature, then the the following option through the command line interface has to be added: {\tt --config use\_bignum=1}.} (i.e. {\tt 1000000000000000000000000001}).
\item {\tt LOOPTYPE=EXPRANGE}. It is similar to {\tt RANGE} but interpolates exponentially between the initial and final elements. Therefore a {\tt STEP=1} means exponential increases by one order of magnitude. Admitted values take the same form than for {\tt RANGE}.
\end{enumerate}
\begin{table}[!ht]
\center
{\tt
\begin{tabular}{ll}
\hline
\textrm{Structure} & \textrm{Assignment} \\
\hline
LOOPTYPE=LIST & VALUE=$r_{i,0}$ $[$VALUE=$r_{i,1}$ $\ldots$ VALUE=$r_{i,m_i-1}]$\\
LOOPTYPE=RANGE & START=$r_{i,0}$ END=$r_{i,m_i-1}$ $[$STEP=$s_{i}$$\mid$POINTS=$m_i$$]$\\
LOOPTYPE=EXPRANGE & START=$r_{i,0}$ END=$r_{i,m_i-1}$ $[$STEP=$e_{i}$$\mid$POINTS=$m_i$$]$\\
\hline
\end{tabular}
}
\caption{Basic syntactic grammar for the $i$-th line of the parameter file. $r_{i,j}$ is the $j$-th value of the set $P_i$, $s_i$ is a linear step increment while $e_i$ defines order of magnitude factor to separate points in the {\tt EXPRANGE}.}
\label{syntax}
\end{table}
A summary of the correct sentence composition is given in \tab{syntax}, where the square brackets ([~]) enclose optional words, and the pipe symbol ($\mid$) refers to mutually exclusive pairs.

Additionally, there are two other keys which may be included for modifying the evaluation of the sets, specially suited to the {\tt RANGE} and {\tt EXPRANGE}:
\begin{itemize}
\item{\tt SKIP}. It permits to remove the specific point from the set.
\item{\tt FUNCTION}. Any valid perl function or concatenation of functions which accept an argument are candidates to transform the $j$-th element in the set $P_i$ according to $r_{i,j} \to f_i(r_{i,j})$.
\end{itemize}
\begin{table}[!ht]
\center
{\tt
\begin{tabular}{lll}
\hline
\textrm{Key=Value} & \textrm{Description}\\
\hline
SKIP=$r_{i,j}$&\\
FUNCTION=$f_i=$ &\\
 abs&{\rm absolute value function}\\
 atan2&{\rm arctangent in the range -$\pi$ to $\pi$}\\
 cos&{\rm cosine function}\\
 exp&{\rm raise to a power}\\
  hex&{\rm convert a string to a hexadecimal number}\\
 int&{\rm get the integer portion of a number}\\
 log&{\rm retrieve the natural logarithm for a number}\\
 oct&{\rm convert a string to an octal number}\\
 rand&{\rm retrieve the next pseudo-random number}\\
 sin&{\rm return the sine of a number}\\
 sqrt&{\rm square root function}\\
 srand&{\rm seed the random number generator}\\
 chomp&{\rm remove a trailing record separator from a string}\\
 chop&{\rm remove the last character from a string}\\
 chr&{\rm get character this number represents}\\
 crypt&{\rm one-way passwd-style encryption}\\
 hex&{\rm convert a string to a hexadecimal number}\\
 lc&{\rm return lower-case version of a string}\\
 lcfirst&{\rm return a string with just the next letter in lower case}\\
 length&{\rm return the number of bytes in a string}\\
 oct&{\rm convert a string to an octal number}\\
 ord&{\rm find a character's numeric representation}\\
 reverse&{\rm flip a string or a list}\\
 uc&{\rm return upper-case version of a string}\\
 ucfirst&{\rm return a string with just the next letter in upper case}\\
\hline
\end{tabular}
}
\caption{Optional tags available for modifying the standard composition of elements in each set.}
\label{optionalsyntax}
\end{table}
A summary of possible option values for the {\tt FUNCTION} key as well as the syntax is given in \tab{optionalsyntax}.
\subsection{Wildcarding}
Once the parsing of the file is finished, the tool has obtained sufficient information to create the Cartesian product given by \eq{cartesian}. A nice feature which allows parameter substitution and job template index substitution is wildcarding. The pattern for substitution has the following syntax:
\begin{itemize}
\item {\tt \$\{1\}, \$\{2\},$\ldots$, \$\{n\}}. These tags are replaced during evaluation by the corresponding values $r_{1}, r_{2}, \ldots, r_{n}$. Note that due to the strict increasing order of replacement, no replacement for {\tt \$\{i\}} is performed at lower set declarations from $P_1$ to $P_{i-1}$.
\item {\tt \$\{JT\_ID\}}. It is replaced by the index ordering the whole set $P$. If $m>10$, so many leading zeroes are included as necessary making file matching and ordering easier.
\end{itemize}
%
\section{Setup}\label{setup}

The Grid[Way] Job Template Manager is a stand-alone perl script, which includes all necessary subroutines for execution.

\subsection{Installation}
A simple download of the program into the current working directory is enough condition to have it installed. 

If the GridWay Metascheduler is missing, only the create and delete operations will be available for use. Instructions to install the Metascheduler are available elsewhere\cite{gridwayinstall} and lay outside the scope of the paper.

\subsection{Configuration}
There are two mechanisms to configure the behaviour of the program.
\begin{itemize}
\item In-line config option. This way allows the user a prompt modification of the configuration parameter through the {\tt --config PARAMETER=VALUE} command line option.
\item Editing the program file. Alternatively, the program file contains all the configuration options ordered in two subroutines Get\_config\_\-parameters and Get\_config\_template\_keywords at the end of the program.
\end{itemize}
\begin{table}
\center
{\tt
\begin{tabular}{ll}
\hline
\textrm{Parameter } &  \textrm{Default Value}\\
\hline
	job\_template\_wildcard & '\$\{JT\_ID\}'\\
	job\_template\_prefix & ''\\
	job\_template\_suffix & '.jt'\\
	std\_output\_dir & '.'\\
	std\_error\_dir & '.'\\
	input\_file\_default\_suffix & '.in'\\
	comment\_char & '\#'\\
	keyassignment\_char & '='\\
	separation\_char & ','\\
	separation\_char\_cli & ' '\\
	separation\_char\_filename & '\_'\\
	jt\_id\_to\_arg\_separation & '\_'\\
    unix\_operators & '\&|<>;()`'\\
	gridway\_submit & 'gwsubmit'\\
	gridway\_ps & 'gwps'\\
	gridway\_kill & 'gwkill'\\
	gridway\_wait & 'gwwait'\\
	gridway\_dir\_var & ''\\
	use\_bignum & 0\\
	huge\_number\_points & 10000\\
	inode\_size\_kB & 4 \\
\hline
\end{tabular}
}
\caption{Get\_config\_keywords parameter available. Strings are enclosed by single ticks.}
\label{keyword}
\end{table}
\begin{table}
\center
{\tt
\begin{tabular}{ll}
\hline
\textrm{Parameter} & \textrm{Default Value}\\
\hline
	Template\_executable & 'EXECUTABLE'\\
	Template\_arguments & 'ARGUMENTS'\\
	Template\_stdout\_file & 'STDOUT\_FILE'\\
	Template\_stderr\_file & 'STDERR\_FILE'\\
	Template\_job\_name & 'NAME'\\
	Template\_encloser\_char & ''\\
	Template\_end\_of\_line & ''\\
\hline
\end{tabular}
}
\caption{Get\_config\_template\_keywords. Strings are enclosed by single ticks.}
\label{templatekeyword}
\end{table}
The whole set of possible parameters to be assigned are given in \tabs{keyword}{templatekeyword}

\section{Usage}\label{usage}

The tool has been designed to be used through the command line interface of the users' terminal. It suffices to invoke the main command together with a subcommand which specifies the action to be performed and arguments to set the scope of the operation. Typed at the prompt, it reads:
\begin{verbatim}
gw_job_template_manager [OPTION] SUBCOMMAND ARG(S)
\end{verbatim}
where some options for debug and configure also exists. All the elements needed for a correct use are summarized in \tab{usagetable}.

\begin{table}[!th]
\center
\begin{tabular}{ll}
\hline
Subcommand & Description\\
\hline
{\tt -c, --create} & create templates\\
{\tt -d, --delete} & delete templates\\
{\tt -s, --submit} & submit the jobs from templates\\
{\tt -p, --purge} & purge the existing jobs from templates\\
{\tt -k, --kill} & kill the existing jobs from templates\\
{\tt -i, --info} & information about status of the submitted jobs\\
{\tt -v, --version} & version number of the program\\
{\tt -l, --license} & credits and license\\
{\tt -h, --help} & print help\\
\hline
Option & Description\\
\hline
{\tt -w, --worker} & add worker file\\
{\tt -t, --template} & add template file\\
{\tt     --signal} & add signal for kill\\
{\tt     --debug} & show debugging information\\
{\tt     --config key=value}& assign a key=value pair for configuration settings\\
\hline
Syntax\\
\hline
\multicolumn{2}{l}{{\tt -c <PARAMETER\_FILE> [-w WORKER\_FILE] [-t TEMPLATE\_FILE]}}\\
\multicolumn{2}{l}{\tt -d <all|[un]submitted|[un]finished|[un]successful|FROM-TO>}\\
\multicolumn{2}{l}{\tt -s <all|[un]submitted|[un]finished|[un]successful|FROM-TO>}\\
\multicolumn{2}{l}{\tt -p <all|[un]finished|[un]successful|FROM-TO>}\\
\multicolumn{2}{l}{\tt -k <all|[un]finished|[un]successful|FROM-TO> [--signal SIG]}\\
\multicolumn{2}{l}{\tt -i <history|now|evolution>}\\
\hline
Argument & Description\\
\hline
{\tt PARAMETER\_FILE} & file with parameter description\\
{\tt WORKER\_FILE} & executable file to be run remotely\\
{\tt TEMPLATE\_FILE} & append extra variables from template file\\
{\tt all} &               all the templates are subcommanded\\
{\tt [un]submitted} &     only those which were [not] submitted\\
{\tt [un]finished} &      only those whose jobs have [not] finished\\
{\tt [un]successful} &    only those whose jobs finished [un]succesfully\\
{\tt FROM-TO} &           deletes the range [FROM,TO]\\
{\tt SIG} &               kill with signal passed to gwkill\\
{\tt history} &           full historic information about each job\\
{\tt now} &               last status update for each job\\
{\tt evolution} &         timely snapshots of job statuses\\
\hline
\end{tabular}
\caption{Different subcommands, syntax, arguments and options for the command line interface. Square brackets indicate optional features whilst angle brackets are compulsory elements.}
\label{usagetable}
\end{table}

Firstly, the user would edit a parameter file containing the instruction to generate the arguments wishing to pass as arguments to the executable. These job templates include five definitions:
\begin{itemize}
\item {\tt EXECUTABLE =} executable filename (i.e. {\tt WORKER\_FILE}),
\item {\tt ARGUMENTS =} argument list, separated by whitespaces,
\item {\tt STDOUT\_FILE =} standard output destination file,
\item {\tt STDERR\_FILE =} standard error destination file,
\item {\tt NAME =} label for naming the job.
\end{itemize}

These definitions correspond to the GridWay Job Template Language and can be extended by adding the {\tt TEMPLATE\_FILE} with extra ones or modified through the config option.

After a successful run of the program, a zero exit status is delivered. Otherwise, it indicates an error, probably associated with either internal code errors or wrong syntax. Because there is no systematic way to represent program exit codes, an indication of the encountered problem is given in \tab{error}.
\begin{table}
\center
\begin{tabular}{cp{50ex}}
\hline
{Exit status} & {Description}\\
\hline
0 & success\\
1 & wrong syntax in the command line\\
2 & system execution error\\
3 & file not found or requirement not matched\\
4 & parameter file syntax inconsistent\\
5 & error opening file\\
6 & error closing file\\
7 & no job found coming from template in the list\\
8 & internal computation error\\
9 & internal parsing error\\
\hline
\end{tabular}
\caption{Success and error code description from exit status.}
\label{error}
\end{table}

\section{Examples}\label{examples}
A simple collection of examples will illustrate the usage of the tool and some tips enhancing the possibilities of writing job templates.

\subsection{Hello world!}
Sticking to tradition, the first example must create a job template whose job is submitted and prints {``Hello world!''}. Doing this involves at least two executions of the Grid[Way] Job Template Manager. Step by step:
\begin{enumerate}
\item A file (say {\tt parameter.in}) describes which is the string argument to be printed. One valid line suffices in this case:
\begin{term}
$ cat parameter.in
LOOPTYPE=LIST, VALUE="Hello world!"
\end{term}
\item Let us run the command\footnote{Note that the command has been preceeded by a {\tt \$} symbol resembling the shell prompt.} with the creation of job templates ({\tt --create}) and an executable of the system ({\tt --worker}):
\begin{term}
$ gw_job_template_manager --create parameter.in --worker /bin/echo
Composed 1 job templates
\end{term}
Once finished, the job template appears in the current directory:
\begin{term}
$ ls
0_echo_Hello_world!.jt parameter.in
\end{term}
with a name indicating the job template index ({\tt 0}), the job executable ({\tt echo}) without trailing path, the sampled argument ({\tt Hello World!}) where the whitespaces have been substituted by underscores, and a suffix extension ({\tt .jt}). It contains five fields with instructions for GridWay:
\begin{term}
$ cat 0_echo_Hello_world!.jt
NAME = 0_echo
EXECUTABLE = /bin/echo
ARGUMENTS = "Hello world!"
STDOUT_FILE = 0_echo_Hello_world!.out
STDERR_FILE = 0_echo_Hello_world!.err
\end{term}
\item The submission step is quite straightforward, since the tag {\tt all} covers the submission of all generated job templates:
\begin{term}
$ gw_job_template_manager --submit all
Submitted 1 jobs from templates
\end{term}
depending on the scheduling, and resource availability, it will take more or less time for the remote job to finish. When it is done, the standard output file will be there containing the expected greeting.
\begin{term}
$ cat  0_echo_Hello_world!.out
Hello world!
\end{term}
\end{enumerate}
\subsection{Interplanetary Hello world!}
As a second non-trivial example, we present an extension of the previous case where the ability to perform a two-dimensional product of sets is shown.
\begin{enumerate}
\item The file {\tt parameter.in} has two lines:
\begin{term}
LOOPTYPE=LIST, VALUE=hello, VALUE=goodbye, FUNCTION=ucfirst
LOOPTYPE=LIST, VALUE=world!, VALUE=mars!
\end{term}
where two new aspects appear. First, we have changed our strategy and instead of echoing a single argument passed as a quoted string, we have given two arguments to the {\tt /bin/echo} worker application. The first of them is composed by a combination of the set $P_1=$\{hello, goodbye\} and the second belongs to the set $P_2=$\{world!, mars!\}. Secondly, there is a  transformation according to the perl function {\tt ucfirst} which capitalizes the first letter of the word in $P_1$.
\item Creating the job templates generates the following files:
\begin{term}
$ ls -1
0_echo_Hello_world!.jt
1_echo_Hello_mars!.jt
2_echo_Goodbye_world!.jt
3_echo_Goodbye_mars!.jt
parameter.in
\end{term} 
with equivalent content as shown in the simple ``Hello world!'' example.
\item After the submission of all the templates, it is very useful to monitor the status of the jobs. This action can be performed with the {\tt --info} subcommand
\begin{term}
$ gw_job_template_manager --info now
JOB_NAME,LOCALTIME,TIME,MANAGER,STATUS,QUEUE_NAME,HOST_NAME,EXIT_STATUS
0_echo,Tue Feb 23 19:24:45 2010,1266949485,DISPATCH,WRAPPER,prod,egee.srce.hr,
1_echo,Tue Feb 23 19:25:05 2010,1266949505,DISPATCH,WRAPPER,gilda,grid.acad.bg,
2_echo,Tue Feb 23 19:24:54 2010,1266949494,EXECUTION,ACTIVE,prod,egee.srce.hr,
3_echo,Tue Feb 23 19:25:04 2010,1266949504,DISPATCH,DONE,prod,egee.srce.hr,0
\end{term}
which outputs a CSV-formatted\footnote{Comma separated values.} table with information about each job. This info subcommand is intended to provide knowledge about the set of jobs of the parameter study, leaving aside those jobs which do not match a parent job template in the current directory.
\item Once the jobs are successfully finished, it is possible to purge them from the list
\begin{term}
$ gw_job_template_manager --purge successful
Purged 4 jobs from templates
\end{term}
and also to get rid of all the job templates
\begin{term}
$ gw_job_template_manager --delete all
Deleted 4 job templates
\end{term}
being in the current directory. Nevertheless, there is some limitation here; the {\tt [un]submitted}, {\tt [un]finished}, and {\tt [un]successful} tags only make sense while matching their respectives jobs from templates listed through the {\tt gwps} GridWay command. This means that if the user purges the successful jobs, only the unsuccessful remain in the list being therefore impossible to pursue more actions (like delete) on a successful set.
\item The output of the jobs now resembles an interplanetary dialog.
\begin{term}
$ cat *.out
Hello world!
Hello mars!
Goodbye world!
Goodbye mars!
\end{term}
\end{enumerate}
\subsection{Squaring and skipping}
The next example tries to compute the square of the set $P=$\{1,2,4,5\}. 
\begin{enumerate}
\item Instead of typing them by hand, a squaring executable  which receives a single argument has been prepared (named {\tt square}). The parameter file {\tt para\-meter.in} has one line:
\begin{term}
LOOPTYPE=RANGE, START=1, END=5, STEP=1, SKIP=3
\end{term}
saying it should fill the integer interval [1,5], and from them skip the value 3
\begin{term}
$ gw_job_template_manager -c parameter.in -w square
Composed 4 job templates
$ ls -1
0_square_1.jt
1_square_2.jt
2_square_4.jt
3_square_5.jt
parameter.in
square
\end{term}
Note that this time, the shortcuts {\tt -c} and {\tt -w} have been use instead of {\tt --create} and {\tt --worker}.
\item Job templates do not need to be submitted all at a time. For instance, the user could perform a first operation for jobs 0 to 1 and, later on, a second operation with the rest.
\begin{term}
$ gw_job_template_manager --submit 0-1 
Submitted 2 jobs from templates
$ gw_job_template_manager --submit unsubmitted
Submitted 2 jobs from templates
\end{term}
\item Another way to get to know what happened to each job is to use the history tag with the information command:
\begin{term}
$ gw_job_template_manager --i history
JOB_NAME,LOCALTIME,TIME,MANAGER,STATUS,QUEUE_NAME,HOST_NAME,EXIT_STATUS
0_square,Wed Feb 24 12:33:35 2010,1267011215,DISPATCH,PENDING,,,
0_square,Wed Feb 24 12:33:38 2010,1267011218,DISPATCH,PROLOG,,,
0_square,Wed Feb 24 12:33:39 2010,1267011219,DISPATCH,WRAPPER,prod,egee.srce.hr,
0_square,Wed Feb 24 12:33:39 2010,1267011219,EXECUTION,FAILED,prod,egee.srce.hr,
0_square,Wed Feb 24 12:33:39 2010,1267011219,DISPATCH,EPILOG_FAIL,prod,egee.srce.hr,
0_square,Wed Feb 24 12:33:39 2010,1267011219,DISPATCH,PENDING,,,
0_square,Wed Feb 24 12:33:48 2010,1267011228,DISPATCH,PROLOG,,,
0_square,Wed Feb 24 12:33:48 2010,1267011228,DISPATCH,WRAPPER,gilda,grid.acad.bg,
0_square,Wed Feb 24 12:33:53 2010,1267011233,EXECUTION,FAILED,gilda,grid.acad.bg,
0_square,Wed Feb 24 12:33:53 2010,1267011233,DISPATCH,EPILOG_FAIL,gilda,grid.acad.bg,
0_square,Wed Feb 24 12:33:53 2010,1267011233,DISPATCH,PENDING,,,
0_square,Wed Feb 24 12:33:58 2010,1267011238,DISPATCH,PROLOG,,,
0_square,Wed Feb 24 12:34:33 2010,1267011273,DISPATCH,WRAPPER,default,gridway.org,
0_square,Wed Feb 24 12:34:33 2010,1267011273,EXECUTION,PENDING,default,gridway.org,
0_square,Wed Feb 24 12:34:38 2010,1267011278,EXECUTION,ACTIVE,default,gridway.org,
0_square,Wed Feb 24 12:34:39 2010,1267011279,EXECUTION,DONE,default,gridway.org,
0_square,Wed Feb 24 12:34:39 2010,1267011279,DISPATCH,EPILOG_STD,default,gridway.org,
0_square,Wed Feb 24 12:34:48 2010,1267011288,DISPATCH,EPILOG,default,gridway.org,
0_square,Wed Feb 24 12:34:57 2010,1267011297,DISPATCH,DONE,default,gridway.org,0
1_square,Wed Feb 24 12:33:35 2010,1267011215,DISPATCH,PENDING,,,
1_square,Wed Feb 24 12:33:39 2010,1267011219,DISPATCH,PROLOG,,,
1_square,Wed Feb 24 12:33:39 2010,1267011219,DISPATCH,WRAPPER,prod,egee.srce.hr,
1_square,Wed Feb 24 12:33:39 2010,1267011219,EXECUTION,FAILED,prod,egee.srce.hr,
1_square,Wed Feb 24 12:33:39 2010,1267011219,DISPATCH,EPILOG_FAIL,prod,egee.srce.hr,
1_square,Wed Feb 24 12:33:39 2010,1267011219,DISPATCH,PENDING,,,
1_square,Wed Feb 24 12:33:48 2010,1267011228,DISPATCH,PROLOG,,,
1_square,Wed Feb 24 12:33:48 2010,1267011228,DISPATCH,WRAPPER,gilda,grid.acad.bg,
1_square,Wed Feb 24 12:33:53 2010,1267011233,EXECUTION,FAILED,gilda,grid.acad.bg,
1_square,Wed Feb 24 12:33:53 2010,1267011233,DISPATCH,EPILOG_FAIL,gilda,grid.acad.bg,
1_square,Wed Feb 24 12:33:53 2010,1267011233,DISPATCH,PENDING,,,
1_square,Wed Feb 24 12:33:58 2010,1267011238,DISPATCH,PROLOG,,,
1_square,Wed Feb 24 12:34:33 2010,1267011273,DISPATCH,WRAPPER,default,gridway.org,
1_square,Wed Feb 24 12:34:33 2010,1267011273,EXECUTION,PENDING,default,gridway.org,
1_square,Wed Feb 24 12:34:38 2010,1267011278,EXECUTION,ACTIVE,default,gridway.org,
1_square,Wed Feb 24 12:34:39 2010,1267011279,EXECUTION,DONE,default,gridway.org,
1_square,Wed Feb 24 12:34:39 2010,1267011279,DISPATCH,EPILOG_STD,default,gridway.org,
1_square,Wed Feb 24 12:34:49 2010,1267011289,DISPATCH,EPILOG,default,gridway.org,
1_square,Wed Feb 24 12:34:58 2010,1267011298,DISPATCH,DONE,default,gridway.org,0
2_square,Wed Feb 24 12:34:01 2010,1267011241,DISPATCH,PENDING,,,
2_square,Wed Feb 24 12:34:08 2010,1267011248,DISPATCH,PROLOG,,,
2_square,Wed Feb 24 12:34:24 2010,1267011264,DISPATCH,WRAPPER,default,gridway.org,
2_square,Wed Feb 24 12:34:24 2010,1267011264,EXECUTION,PENDING,default,gridway.org,
2_square,Wed Feb 24 12:34:33 2010,1267011273,EXECUTION,ACTIVE,default,gridway.org,
2_square,Wed Feb 24 12:34:34 2010,1267011274,EXECUTION,DONE,default,gridway.org,
2_square,Wed Feb 24 12:34:34 2010,1267011274,DISPATCH,EPILOG_STD,default,gridway.org,
2_square,Wed Feb 24 12:34:41 2010,1267011281,DISPATCH,EPILOG,default,gridway.org,
2_square,Wed Feb 24 12:34:56 2010,1267011296,DISPATCH,DONE,default,gridway.org,0
3_square,Wed Feb 24 12:34:01 2010,1267011241,DISPATCH,PENDING,,,
3_square,Wed Feb 24 12:34:08 2010,1267011248,DISPATCH,PROLOG,,,
3_square,Wed Feb 24 12:34:22 2010,1267011262,DISPATCH,WRAPPER,default,gridway.org,
3_square,Wed Feb 24 12:34:24 2010,1267011264,EXECUTION,PENDING,default,gridway.org,
3_square,Wed Feb 24 12:34:34 2010,1267011274,EXECUTION,ACTIVE,default,gridway.org,
3_square,Wed Feb 24 12:34:34 2010,1267011274,EXECUTION,DONE,default,gridway.org,
3_square,Wed Feb 24 12:34:34 2010,1267011274,DISPATCH,EPILOG_STD,default,gridway.org,
3_square,Wed Feb 24 12:34:41 2010,1267011281,DISPATCH,EPILOG,default,gridway.org,
3_square,Wed Feb 24 12:34:55 2010,1267011295,DISPATCH,DONE,default,gridway.org,0
\end{term}
where the example shows how jobs 0 and 1 failed after trying to be run without permissions at different resources and ended up in another resource where jobs were allowed to run. The GridWay Metascheduler learned from that and managed jobs 2 and 3 more efficiently.
\item The output shows a properly squared set
\begin{term}
$ cat  *.out
1^2=1
2^2=4
4^2=16
5^2=25
\end{term}
\end{enumerate}
\subsection{Random Monte-Carlo}
Often, programs involving large statistical runs require a proper initiation of the random number generator. This can be done by means of a seed and is a key point for ensuring uncorrelated jobs. Here a simple way of getting different seeds in the integer interval [0,1000) is implemented for eight independent jobs.
\begin{enumerate}
\item The {\tt parameter.in} looks like:
\begin{term}
$ cat parameter.in
LOOPTYPE=RANGE, START=1000, END=1000, POINTS=8, \
FUNCTION=int rand
\end{term}
and note that evaluating eight jobs within 1000 and 1000 makes an apparently dummy set $\{1000, \ldots ,1000\}$, but the transformation under the function composition {\tt int rand} generates eight random numbers and throws away the decimal part.
\item A possible output of the creation
\begin{term}
$ gw_job_template_manager -w worker -c parameter.in
Composed 8 job templates
\end{term}
is
\begin{term}
$ ls -1
0_worker_292.jt
1_worker_741.jt
2_worker_468.jt
3_worker_481.jt
4_worker_159.jt
5_worker_310.jt
6_worker_555.jt
7_worker_645.jt
parameter.in
worker
\end{term}
\end{enumerate}
\subsection{Advanced wildcarding}
Another common way to produce results is writing to an output file instead of to the standard output. In such a case, let us consider a worker executable comparing a Taylor expansion and an exact result for several orders of magnitude. It takes three arguments, being the last one a string containing the file name in which results are to be written.
\begin{enumerate}
\item The parameter file could be
\begin{term}
$ cat parameter.in
LOOPTYPE=LIST, VALUE=Taylor, VALUE=Exact
LOOPTYPE=EXPRANGE, START=1, END=1E3, STEP=1, SKIP=100
LOOPTYPE=LIST, VALUE=${JT_ID}.txt
\end{term}
with three features not yet seen: the {\tt EXPRANGE} type which ranges exponentially, the scientific notation {\tt 1E3} and a last line, indicating that part of the third argument is going to be translated into the job template index {\tt \${JT\_ID}} for each job.
\item The creation of templates is this time slightly different due to the retrieval of extra output files. For this to be taken into account, an extra job template input file is needed {\tt template.in}:
\begin{term}
$ cat template.in
OUTPUT_FILES=${3}
$ gw_job_template_manager -w worker -c parameter.in -t worker.in
Composed 6 job templates
\end{term}
and the result
\begin{term}
$ ls -1
0_worker_Taylor_1_0.txt.jt
1_worker_Taylor_10_1.txt.jt
2_worker_Taylor_1000_2.txt.jt
3_worker_Exact_1_3.txt.jt
4_worker_Exact_10_4.txt.jt
5_worker_Exact_1000_5.txt.jt
parameter.in
worker
template.in
$ cat 5_worker_Exact_1000_5.txt.jt
NAME = 5_worker
EXECUTABLE = worker
ARGUMENTS = Exact 1000 5.txt
STDOUT_FILE = 5_worker_Exact_1000_5.txt.out
STDERR_FILE = 5_worker_Exact_1000_5.txt.err
OUTPUT_FILES=5.txt
\end{term}
\item The submission does not require any difference, but this time we want to know how much do the jobs take to finish:
\begin{term}
$ gw_job_template_manager -s all; time gw_job_template_manager -p all
Submitted 6 jobs from templates
Purged 6 jobs from templates

real    1m51.861s
user    0m0.660s
sys	    0m0.060s
\end{term}
\item In this case, the relevant results are the text files
\begin{term}
$ ls -1 *.txt
0.txt
1.txt
2.txt
3.txt
4.txt
5.txt
\end{term}
\end{enumerate}
\subsection{Big numbers}
What if the input are very large numbers? Next illustrates how to proceed
\begin{enumerate}
\item Let consider the case in which the parameter file is
\begin{term}
$ cat parameter.in
LOOPTYPE=RANGE,\
START=123456789012345678911234567892123456789312345678941,\
  END=123456789012345678911234567892123456789312345678943,\
POINTS=3
\end{term}
\item The standard tentative will {\bf not} work\footnote{Depending on the perl version, for perl 5.10 it does not work, but earlier versions (5.8) did not yet transform automatically into scientific notation.}
\begin{term}
$ gw_job_template_manager -w /bin/echo -c parameter.in
Composed 3 job templates
$ ls -1
0_echo_1.23456789012346e+50.jt
1_echo_1.23456789012346e+50.jt
2_echo_1.23456789012346e+50.jt
parameter.in
\end{term}
because the transformation into scientific notation dropped out the precision we expect to work with. 
\item Instead, the user should instruct the tool to profit from the {\tt bignum} perl package
\begin{term}
$ gw_job_template_manager --delete all
Deleted 3 job templates
$ gw_job_template_manager --config use_bignum=1\
  -w /bin/echo -c parameter.in
Composed 3 job templates
$ ls -1
0_echo_123456789012345678911234567892123456789312345678941.jt
1_echo_123456789012345678911234567892123456789312345678942.jt
2_echo_123456789012345678911234567892123456789312345678943.jt
parameter.in
\end{term}
so obtaining the desired job templates.
\end{enumerate}
\subsection{Using gLite Job Description Language}
When using the middleware gLite, the user will probably find a user interface without the GridWay Metascheduler installed. In this example a method to translate the job templates into JDL and how to submit them is described.
\begin{enumerate}
\item Four simple tests are prepared. They give a hint about the working environment:
\begin{term}
$ cat parameter.in
LOOPTYPE=LIST, VALUE=ps, VALUE=pwd, VALUE=ls, VALUE=whoami
$ cat template.in
OutputSandbox = {"${JT_ID}_env_${1}.out","${JT_ID}_env_${1}.err"};
\end{term}
\item A way to set up the propper grammar in the template file is to change the keys and some formating elements through config options:
\begin{term}
$ gw_job_template_manager \
-w /usr/bin/env -c parameter.in -t template.in \
--config Template_executable=Executable \
--config Template_arguments=Arguments \
--config Template_stdout_file=StdOutput \
--config Template_stderr_file=StdError \
--config Template_job_name=JobName \
--config Template_encloser_char=\" \
--config Template_end_of_line=\; \
--config job_template_suffix=.jdl
WARNING: GridWay location not set up.
This means that the usability of this tool is limited to create and delete 
job templates. Please identify your $GW_LOCATION directory and set the 
parameter to that value with "--config GW_LOCATION=value".
Composed 4 job templates
\end{term}
\item The Grid[Way] Job Template Manager does not know how to submit templates for other middleware, but the user can still tweak the settings through the {\tt --config} modifier and get the submission procedure done:
\begin{term}
$ gw_job_template_manager -s all \
--config gridway_submit=glite-wms-job-submit \
--config gridway_submit_flag=-a \
--config job_template_suffix=.jdl
WARNING: GridWay location not set up.
This means that the usability of this tool is limited to create and delete 
job templates. Please identify your $GW_LOCATION directory and set the 
parameter to that value with "--config GW_LOCATION=value".

Connecting to the service https://gilda-rb.rediris.es:7443/glite_wms_wmproxy_server


====================== glite-wms-job-submit Success ======================

The job has been successfully submitted to the WMProxy
Your job identifier is:

https://gilda-rb.rediris.es:9000/9xFbq28QPgTcYLxCeVU5Kw

==========================================================================



Connecting to the service https://gilda-rb.rediris.es:7443/glite_wms_wmproxy_server


====================== glite-wms-job-submit Success ======================

The job has been successfully submitted to the WMProxy
Your job identifier is:

https://gilda-rb.rediris.es:9000/IP4LKZgkpxcL9NE58AHuPQ

==========================================================================



Connecting to the service https://gilda-rb.rediris.es:7443/glite_wms_wmproxy_server


====================== glite-wms-job-submit Success ======================

The job has been successfully submitted to the WMProxy
Your job identifier is:

https://gilda-rb.rediris.es:9000/3fyT-LnV3xkimEPfSsAHag

==========================================================================



Connecting to the service https://gilda-rb.rediris.es:7443/glite_wms_wmproxy_server


====================== glite-wms-job-submit Success ======================

The job has been successfully submitted to the WMProxy
Your job identifier is:

https://gilda-rb.rediris.es:9000/53HM4HRBCsqwHwpa01zEVw

==========================================================================


Submitted 4 jobs from templates
\end{term}
\item Finally, by picking up the job identifiers of the output, a final recovery of data is straightforward but not possible with the help of the tool. For example, to retrieve the output of the last job:
\begin{term}
$ glite-wms-job-output https://gilda-rb.rediris.es:9000/IP4LKZgkpxcL9NE58AHuPQ
Connecting to the service https://gilda-rb.rediris.es:7443/glite_wms_wmproxy_server


================================================================================

			JOB GET OUTPUT OUTCOME

Output sandbox files for the job:
https://gilda-rb.rediris.es:9000/53HM4HRBCsqwHwpa01zEVw
have been successfully retrieved and stored in the directory:
/tmp/jobOutput/user_53HM4HRBCsqwHwpa01zEVw

================================================================================
$ cd /tmp/jobOutput/user_53HM4HRBCsqwHwpa01zEVw
$ ls -1
3_env_whoami.err
3_env_whoami.out
$ cat 3_env_whoami.out
gilda075
\end{term}
\end{enumerate}

\section{Conclusion}\label{conclusion}
A tool for managing a parameter sweep study on a distributed architecture has been presented, with special emphasis on the mechanism to create job templates from a given parameter file. The syntax of the parameter file is simple though quite powerful, specially because broad scoped features (arbitrary dimensions, exponentiation, value skipping, wildcarding and functions) have been implemented.

The usage has been designed to perform an action and the anti-action, so to the creation of job templates a deletion exists. Moreover, submission of jobs is also complemented by the purge. In any case, all the operations are grouped under useful subcommands which can be accessed through the command line interface.

In order to allow enough flexibility, the user has at disposal configuration options to modify program parameters. Some examples have been worked out for illustrating how a little amount of input give rise to a plethora of possibilities, from the ``Hello world!'' trivial case up to a  parameter study controlled with the GridWay Metascheduler.

Within the aspects which could be desirable but not yet implemented, it can be mentioned the capability to handle job templates in the XML-formatted Job Standard Description Language. The problem to accept this format is the inherent hierarchy of the XML syntax, which is incompatible with appending a template file with extra information. Nevertheless, an smart merge might be implemented under certain scheme. Another kind of limitation is the use of wildcards and functions altogether, or the need of other wildcards. The inclusion of these features could be studied if they turn to handicap parameter sweep studies from the scientific community.

The tool can be viewed an extension of GridWay, which performs the resource discovering, scheduling or workloading for the jobs. Being modular, it is much easier to focus on the aspects regarding parameter sweeps and the programming protocol to interact with the middleware. It has been shown also the possibilities to interact with other submission agents, such as the gLite WMS.

Altogether, this tool provides the user a powerful mechanism to port applications to clusters, grid or cloud. Our experience on this subject indicates that the steep learning curve and lack of success are the main drawbacks for adopting the solutions of distributed computing. Offering an opportunity to save time on the generation of the parameters to be swept and to ensure consistent operations on specific blocks of jobs is clearly an step forward to enhance the overall performance of the application at the user's convenience.

\section*{Acknowledgements}
We would like to thank the gilda VO \cite{gilda} for providing a working infrastructure for tests within the EGEE project \cite{EGEE}. Also, we would like to express our gratitude to Antonio Fuentes and Virginia Martín-Rubio from RedIRIS for the kind organization of several tutorials for grid application developers, giving us the opportunity to interact with interested groups of students and researchers. They all gave us unvaluable feedback to improve the tool. 

This work was supported by the Consejería de Educación de la Comunidad de Madrid, Fondo Europeo de Desarrollo Regional (FEDER) and Fondo Social Europeo (FSE), through the MEDIANET Research Program S2009/TIC-1468, by the spanish Ministerio de Ciencia e Innovacion, through the research grant TIN2009-07146, and by the European Union through the EGEE-III grant agreement INFSO-RI-222667.
 

\end{document}